\documentclass[a4paper,12pt, epsfig]{article}
\usepackage{epsfig}
\usepackage{amssymb}
\usepackage{amsfonts}
\usepackage{amsmath}

\newskip\humongous \humongous=0pt plus 1000pt minus 1000pt

\newif\ifdtup

\catcode`\@=11

\@addtoreset{equation}{section}
\def\theequation{\thesection.\arabic{equation}}

\def\@normalsize{\@setsize\normalsize{15pt}\xiipt\@xiipt
\abovedisplayskip 14pt plus3pt minus3pt%
\belowdisplayskip \abovedisplayskip
\abovedisplayshortskip \z@ plus3pt%
\belowdisplayshortskip 7pt plus3.5pt minus0pt}

\def\small{\@setsize\small{13.6pt}\xipt\@xipt
\abovedisplayskip 13pt plus3pt minus3pt%
\belowdisplayskip \abovedisplayskip
\abovedisplayshortskip \z@ plus3pt%
\belowdisplayshortskip 7pt plus3.5pt minus0pt
\def\@listi{\parsep 4.5pt plus 2pt minus 1pt
     \itemsep \parsep
     \topsep 9pt plus 3pt minus 3pt}}

\relax

\catcode`@=12

\catcode`\@=11

\def\section{\@startsection{section}{1}{\z@}{3.5ex plus 1ex minus
   .2ex}{2.3ex plus .2ex}{\large\bf}}

\def\thesection{\arabic{section}}
\def\thesubsection{\arabic{section}.\arabic{subsection}}

\def\appendix{\setcounter{section}{0}
 \def\thesection{Appendix \Alph{section}}
 \def\thesubsection{\Alph{section}.\arabic{subsection}}
 \def\theequation{\Alph{section}.\arabic{equation}}}

\def\SymBoxes#1#2#3#4{\newdimen\un@t \un@t#3%
\raisebox{#1}{\rule{#2\un@t}{#4}\hskip-#2\un@t
\@tempdimb\un@t \advance\@tempdimb by-#4\@tempcntb#2\relax%
\@whilenum{\@tempcntb>0}\do{
\rule{#4}{\un@t}\hskip\@tempdimb \advance\@tempcntb by\m@ne}%
\hskip-#2\un@t \rule[\un@t]{#2\un@t}{#4}%
\rule[\un@t]{#4}{#4}\hskip-#4
\rule{#4}{\un@t}}\hskip-#4}                

\begin{document}

\newcommand{\beq}{\begin{equation}}
\newcommand{\eeq}{\end{equation}}
\newcommand{\bea}{\begin{eqnarray}}
\newcommand{\eea}{\end{eqnarray}}
\newcommand{\beas}{\begin{eqnarray*}}
\newcommand{\eeas}{\end{eqnarray*}}
\newcommand{\defi}{\stackrel{\rm def}{=}}
\newcommand{\non}{\nonumber}
\newcommand{\bquo}{\begin{quote}}
\newcommand{\enqu}{\end{quote}}
\renewcommand{\(}{\begin{equation}}
\renewcommand{\)}{\end{equation}}
\def\IZ{{\mathbb Z}}
\def\IR{{\mathbb R}}
\def\IC{{\mathbb C}}
\def\IQ{{\mathbb Q}}

\newcommand{\nn}{\nonumber}

\def\CM{{\mathcal{M}}}
\def\dCM{{\left \vert\mathcal{M}\right\vert}}

\def \d{\textrm{d}}
\def \p{\partial}

\def\tphi{\tilde\phi}
\def\txi{\tilde\xi}

\def \Pf{\rm Pf\ }

\def \pr{\prime}

\def \M{\mathcal M}

\def\Tr{ \hbox{\rm Tr}}
\def\half{\frac{1}{2}}

\def \eqn#1#2{\begin{equation}#2\label{#1}\end{equation}}
\def\de{\partial}
\def\Tr{ \hbox{\rm Tr}}
\def\H{ \hbox{\rm H}}
\def\HE{ \hbox{$\rm H^{even}$}}
\def\HO{ \hbox{$\rm H^{odd}$}}
\def\K{ \hbox{\rm K}}
\def\Im{ \hbox{\rm Im}}
\def\Ker{ \hbox{\rm Ker}}
\def\const{\hbox {\rm const.}}
\def\o{\over}
\def\im{\hbox{\rm Im}}
\def\re{\hbox{\rm Re}}
\def\bra{\langle}\def\ket{\rangle}
\def\Arg{\hbox {\rm Arg}}
\def\Re{\hbox {\rm Re}}
\def\Im{\hbox {\rm Im}}
\def\exo{\hbox {\rm exp}}
\def\diag{\hbox{\rm diag}}
\def\longvert{{\rule[-2mm]{0.1mm}{7mm}}\,}
\def\a{{\textsl a}}
\def\dag{{}^{\dagger}}
\def\tq{{\widetilde q}}
\def\p{{}^{\prime}}
\def\W{W}
\def\N{{\cal N}}
\def\hsp{,\hspace{.7cm}}
\newcommand{\C}{\ensuremath{\mathbb C}}
\newcommand{\Z}{\ensuremath{\mathbb Z}}
\newcommand{\R}{\ensuremath{\mathbb R}}
\newcommand{\rp}{\ensuremath{\mathbb {RP}}}
\newcommand{\cp}{\ensuremath{\mathbb {CP}}}
\newcommand{\vac}{\ensuremath{|0\rangle}}
\newcommand{\vact}{\ensuremath{|00\rangle}}
\newcommand{\oc}{\ensuremath{\overline{c}}}

\def\Dt{$\rm D3$}
\def\aDt{$\rm{\overline D}3$}

\def\M{\mathcal{M}}
\def\F{\mathcal{F}}
\def\d{\textrm{d}}

\def\Ra{R_{\rm{\overline D}3}}

\def\eps{\epsilon}

\def\N{N_{\rm D3}}
\def\aN{{\bar N}_{\rm{\overline  D}3}}

\begin{flushright}
\end{flushright}

\vspace{18pt}
\begin{center}
{\Large \textbf{On gravity dual of a metastable vacuum in Klebanov-Strassler theory}}
\end{center}

\vspace{6pt}
\begin{center}
{\large\textsl{Anatoly Dymarsky}\\}
\vspace{25pt}
\textit{\small School of Natural Sciences, Institute for Advanced Study,\\Princeton, NJ, 08540}\\ \vspace{6pt}

\end{center}

\vspace{12pt}
We discuss a supergravity description of the metastable state that is created by a stack of \aDt-branes placed at the tip of the KS background.
When the number $p$ of the \aDt-branes is large $g_s p\gg 1$ the characteristic curvature of the corresponding gravity dual
is large in stringy units and one may expect the background to be regular everywhere. Starting from the distances of order $R \sim (g_s p )^{1/4}\alpha'^{1/2}$ away from the tip the new background can be well approximated by a linear perturbation around KS. By applying the appropriate boundary conditions in both IR and UV we found the lowest KK mode of the corresponding linear perturbation.
The solution we found contains VEVs of the $SU(2)\times SU(2)$ invariant operators at the linear order in $p$. As a non-trivial check we calculate
the ADM mass which exactly matches the probe approximation. As a byproduct we also found a gravity background dual to the KS
theory deformed by the operators $W^2$ and $W^2\bar{W}^2$ with small coefficients.

\vspace{4pt} {\small

\noindent }

\newpage

\vspace{1cm}

\renewcommand{\thefootnote}{\arabic{footnote}}
\section{Introduction and Summary}
The SUSY-breaking metastable states \cite{ISS} is an interesting topic which plays important role in different phenomenological applications. One famous example is the metastable state in $SU(M(k+1)-p)\times SU(Mk-p)$ theory
found by Kachru, Pearson, and Verlinde \cite{KPV}. They start with the $SU(M(k+1))\times SU(Mk)$ theory dual to the deformed conifold solution of Klebanov and Strassler \cite{KS}
and place $p$ \aDt-branes at the tip of the conifold. As a result of non-abelian dynamics the stack of \aDt-branes blow into a NS5-brane. The NS5-brane carries $p$ units of \aDt-brane charge. The corresponding SUSY-breaking state turns out to be metastable when $p/M$ is smaller than a certain critical value of order $0.08$.  This state can tunnel into the true vacuum with NS5-brane disappearing and $M-p$ \Dt-branes emerging. Although the field theory description of the metastable vacuum is out of control due to strong coupling, the schematic picture is the following. The original $SU(M(k+1))\times SU(Mk)$ theory has a rich space of vacua which includes the baryonic branch, described by the KS solution, and the mesonic branches labeled by $l\geq 1$
described by the KS solution with $l M$ mobile \Dt-branes placed on the conifold. After placing additional $p$ \aDt-branes the mesonic branches survive while the baryonic branch gets uplifted. Thus the metastable state of the $SU(M(k+1)-p)\times SU(Mk-p)$  theory is the former baryonic branch of the $SU(M(k+1))\times SU(Mk)$ uplifted in the presence of the \aDt-branes.

Since the field theory does not provide a reliable description of the metastable state it is desirable to describe the corresponding physics with help of the gravity dual background.
A priori we do not know if supergravity approximation will be reliable near the location of \aDt-branes, but in any case sufficiently far from the tip the supergravity description should be reliable. Therefore, no matter how complicated the dynamics is in the direct vicinity of the \aDt-branes, in the end it should boil down to some boundary conditions for the supergravity fields at the location of \aDt-branes. These boundary conditions come from the probe action for the \aDt-branes for $p\sim 1$ or NS5-brane for $p\gg 1$. To illustrate why the probe action leads to the correct boundary conditions despite possible complicated dynamics we turn to the classical electrodynamics. The point charge $q$ creates a potential $A_0={q/r}$ that is singular at the origin. The classical electrodynamics breaks down at the distances shorter than the inverse electron mass $m_e^{-1}$. One has to rely on QED beyond that scale. Yet classical electrodynamics is perfectly valid for the distances much larger than $m_e^{-1}$. Moreover to find the correct boundary conditions for $A_0$ at the origin one does not have to know QED. Rather the boundary conditions come from the probe action $\int d^4 x A_\mu j_\mu$ that can be established by probing physics at small energies i.e. distances larger than $m_{e}^{-1}$ away from the charge. The story with the \aDt-branes is no different.

Although the task of finding the supergravity solution dual to the metastable state is well-defined, technically it is very challenging. First few steps were taken in the original paper \cite{KPV}
and later in \cite{dWKV}. The qualitative picture is the following. Very close to the \aDt-branes, at the scales much smaller than the curvature scale of the
KS solution $R_{KS}^2\sim (g_s M)^{1/2}\alpha'$ the geometry looks like an $AdS_5$ throat of radius $\Ra^2\sim (g_s p)^{1/2}\alpha'$ created by the \aDt-branes. If $p\sim 1$ the corresponding geometry is highly curved and no supergravity description valid everywhere is available. But when $g_s M\gg g_s p\gg 1$ the curvature radius $\Ra$ is large in the stringy units and the solution is expected to be regular everywhere. The $AdS_5$ throat created by the \aDt-branes is corrected by small perturbations coming from the KS backround. Some of these perturbations break (anti)SUSY of the \aDt-branes. There are also modes, such as the ISD flux, which are relevant in the dual field theory sense. The relevant modes increase as we come closer to NS5-brane and may, though not necessarily, destroy the geometry. If we focus on the effect of the ISD flux and neglect the effect of other modes the deep IR region near the NS5-brane will look similar to the Polchinski-Strassler solution \cite{PS}. In fact the latter is exactly the IR region of the $AdS_5$ created by the NS5-brane carrying the \Dt-charge, perturbed by the relevant AISD mode. Hence we expect the full supergravity background dual to the metastable state to schematically look as follows. In the very IR near the NS5-brane the solution will look somewhat similar to the (anti)Polchinski-Strassler solution. As we move away from the NS5-brane the Polchinski-Strassler solution
turns into the $AdS_5$ throat perturbed by some relevant modes with small coefficients. The $AdS_5$ is glued to the tip of the Klebanov-Strassler solution forming a smooth weakly curved background. Clearly to find such a solution beyond our reach at this point. Nevertheless we may not need the full solution to describe interesting physics. If we are interested in the vevs of various operators in the metastable state or in numerous phenomenological applications of SUSY-breaking,
it is enough to know the behavior of the solution at large radii $r\gg \Ra$. In this regime the solution in question is only slightly different from the original KS background and can be found in linear expansion around the latter. In the linear approximation we will loose quadratic and higher in $p$ effects e.g. the difference between the stack of \aDt-branes and the NS5-brane. Yet
this approximation contains all leading in $p$ effects e.g.  SUSY breaking. Moreover working at the linear level admits further simplifications: it is self-consistent to focus only on the $SU(2)\times SU(2)$ invariant modes. Let us point out that such a solution is a part of a more complicated linearized solution produced by the localized \aDt-branes at the tip, not a solution produced by the \aDt-branes smeared around the tip. The smeared configuration is not stable as the  \aDt-branes attract each other.

Up to date there were several works attempting to construct the $SU(2)\times SU(2)$ invariant part of the linearized solution. The pioneering work \cite{dWKM} used the expansion around the KT solution \cite{KT} to identify the modes produced by the \aDt-branes. Since the KT solution is singular at the origin, the proper boundary conditions in the IR could not be imposed and
the resulting solution was valid only up to the leading order in $1/r^4$. Although this accuracy is enough for many interesting applications it is not enough, for example, to determine the force induced on a probe \Dt-brane placed in the background. Another work \cite{mGSS} attempted to find a solution expanding in Taylor series  near the tip. Such a solution can not be continued to the UV region. As a result half of boundary conditions can not be imposed and the resulting solution can not be identified. The omission of the approaches based on the perturbative expansion in small or large radius was intended to be cured by the approach taken by Bena, $\rm Gra\tilde{n}a$, and Halmagyi in \cite{Bena:2009xk} (see also \cite{Bena:2010ze} and \cite{Bena:2011hz} for further development).
They used the technique developed by  Borokhov  and Gubser in \cite{BG} and latter refined in \cite{KuSo}. The idea is to make a change of variables representing seven second order equations governing  linear perturbations around the KS background through fourteen first order equations. This allows to find the solutions valid at any radius, albeit in an indirect form.
The authors of \cite{Bena:2009xk} found that the boundary conditions for the stack of \aDt-branes will necessity require the solution to be singular at the tip. While the singularities of the warp factor
and the RR 4-form directly follow from the boundary conditions provided by the \aDt-branes and were expected, the singularity of the 3-form flux was confusing. Indeed the seven modes system describing linearized deformations of the KS background has fourteen solutions: seven should be regular in the IR while singular in the UV and seven should be regular in the UV while regular in the IR. This logic suggests that one should be able to pick any boundary conditions in the IR at will, at least if the UV behavior is not restricted. Yet it was emphasized in \cite{Bena:2009xk} that once the solution contains the singular mode directly related to the presence of the \aDt-branes the 3-form flux becomes singular as well. The same observation was also made in \cite{mGSS}. Since the \aDt-branes directly source only one IR singular mode (the combination of the warp factor and the RR 4-form $\Phi_-$ in the notations of \cite{gkp}) that  does not include any modes related to the 3-form flux, why then one can not  choose the 3-form flux to be regular?
This is due to the coupling of $\Phi_-$ to the flux which is evident from the following equation of motion \cite{gkp}
\bea
\label{eqflux}
(d+i{d\tau\over \Im \tau}\wedge \Re)(\Phi_-G_++\Phi_+G_-)=0\ .
\eea
Here $G_\pm$ are the ISD and IASD parts of the 3-flux while $\Phi_+$ is yet another combination of the warp-factor and the RR 4-form. $\Phi_+$ couples to the \Dt-branes and stays regular in their absence.
In the KS case $\Phi_-=G_-=0$ while $\Phi_+,G_+\neq 0$. After introducing the \aDt-branes, $\Phi_-$ calculated at linear order diverges at the origin.
Hence to compensate this divergence in (\ref{eqflux}) there should be a singular flux $G_-$. Let us note here that at the non-linear level very close to the \aDt-branes $\Phi_-$ will be proportional to the inverse warp-factor $\Phi_-\sim h^{-1}$ and hence regular. That is why the reason for the IASD flux to be singular will disappear in a full non-linear solution.

The aforementioned singular mode of the 3-form flux is in fact an IR regular mode of the linearized system  \cite{Berg:2006xy}. In particular it is normalizable near the tip. There is another mode of the 3-form flux which is more singular at the tip but is UV regular. This mode may not and should not be turned on in the solution describing the \aDt-branes placed at the tip. Here we outlined the behavior of the solution near the tip only schematically. Later in the text we will return to this point.

As a side note we would like to emphasize that the singularity of the 3-form flux discussed above is not an effect of smearing. The $SU(2)\times SU(2)$ invariant linearized solution is
the lowest KK mode of the linearized solution sourced by the localized \aDt-branes. Therefore if the former is singular so is the latter. This is also evident from (\ref{eqflux}) as the linearized $\Phi_-$ is singular in the localized case as well. Rather the singularity of the 3-form flux is an artifact of linearization. In any way we can only trust linear solution to describe physics for the distances $r\gg (g_s p)^{1/4}{\alpha'}^{1/2}$ away from the tip, where the solution is regular.

In this note we construct the $SU(2)\times SU(2)$ invariant mode of the linearized solution describing the metastable state of $SU(M(k+1)-p)\times SU(Mk-p)$ theory in the leading in $p$ order.
We further develop the approach of \cite{Bena:2009xk} and show that there is a unique solution that satisfies the proper boundary conditions in the IR and the UV.
We found this solution expressed indirectly in terms of some integrals. We use this solution to calculate the ADM mass of the metastable  state.
It turns out to be equal to the probe value ${2pT_3  h^{-1}_{KS}(0)}$. As the ADM mass is sensitive to the IR boundary conditions this provides a non-trivial check of our result.
As a by-product we also obtain a gravity dual of the KS theory perturbed by the  operators $W^\alpha W_\alpha$ and $W^\alpha W_\alpha \bar{W}^\beta \bar{W}_\beta$ with small coefficients. The resulting solutions are regular everywhere and can be used to study gauge theories with softly broken SUSY.

This paper is organized as follows. In Section 2 we review the system of  linear equations governing the perturbations around the KS background and find a general solution parametrized by 14 integration constants. In Section 3 we formulate the boundary conditions for
\Dt\ and \aDt-branes and construct the $SU(2)\times SU(2)$ invariant mode of the linearized solution describing the metastable state. In Section 4 we calculate the ADM mass of the metastable state. In Section 5 we briefly describe the gravity background dual to the KS theory perturbed at linear level
by the $SU(2)\times SU(2)$ invariant, parity and ${\mathcal I}$-symmetry even operators.

\section{System of Linearized Equations}
We start with a system of coupled ODEs introduced in \cite{BG} and follow the notations of \cite{Bena:2009xk}.
The system is specified by the kinetic term
\bea
\label{fieldmetric}
&&G_{ab} d\phi^{a} d\phi^{b}= \\ &&e^{4p + 4A} \left(dx^2+ \frac12 dy^2+6dp^2-6 dA^2+
\frac14 e^{-\Phi-2x} (e^{-2y} df^2+e^{2y} dk^2+ 2 e^{2\Phi} dF^2 )+ \tfrac14 d\Phi'^2 \right) \ ,
\nonumber
\eea
and the superpotnetial
\bea
\label{superpotential}
W(\phi)=e^{4A-2p-2x}+e^{4A+4p} \cosh y + \frac12 e^{4A+4p-2x} \left( f \, (2P-F)+kF \right)\ .
\eea
The solution can be expanded $\phi=\phi_0+\delta\phi+..$
where $\phi_0$ solves the unperturbed system
\bea
\label{susyeq}
\left. {d\phi_0^a\over dt}=-{G^{ab}\over 2}{\partial W\over \partial \phi^b}\right|_{\phi=\phi_0}\ .
\eea
The equations for $\delta\phi$ could be linearized
\bea
\frac{d\xi_a}{d t} &=&  \xi_b M^b{}_a\  ,  \label{xieq} \\
\frac{d\delta \phi^a}{d t}  &=& - M^a{}_b \delta\phi^b - G^{ab} \xi_b \label{phieq} \ ,
\eea
where
\bea
\left.\qquad M^a{}_b\equiv\frac12 \frac{\partial}{\partial \phi^b} \left( G^{ac} \frac{\partial W}{\partial \phi^c} \right)\right|_{\phi=\phi_0} \ .
\eea

The unperturbed solution describing the KS background  (we put $g_s$ and the deformation parameter $\epsilon$ to be equal to one) is
\bea \label{KSbackground}
 e^{x_0}&=& \frac14 \, h^{1/2}(\cosh t \sinh t  - t)^{1/3} \ , \nn \\
 e^{y_0}&=&\tanh(t/2) \ , \nn \\
 e^{6 p_0}&=&  \, \frac{  24 \, (\cosh t \sinh t - t)^{1/3}}{ h \, \sinh^2 t}  \ , \nn \\
 e^{6A_0}&=&\frac{1}{3 \cdot 2^9} \, h (\cosh t \sinh t - t)^{2/3} \sinh^2 t   \ , \\
 f_0&=&-P\frac{  \, (t \coth t -1)(\cosh t -1)}{\sinh t}, \nn\\
 k_0&=&-P \frac{\,(t \coth t -1)(\cosh t +1)}{\sinh t}, \nn \\
  F_0&=& P\frac{(\sinh t -t)}{\sinh t} ,\nn \\
  \Phi_0&=&0\nn\ .
\eea
Here $P=-{1\over 4}M\alpha'$ and the warp-factor is defined as
\bea
 h&=&e^{-4A_0-4p_0+2x_0} \non \\
 &=&h_0-32 P^2 \int_0^\tau \frac{t \coth t -1}{\sinh^2t} (\cosh t \sinh t - t)^{1/3} dt \ . \label{KSh}\nn
\eea
The constant $h_0=h(0)$ is chosen such that $h$ approaches zero at infinity.

To write down the linearized equations for $\delta\phi,\xi$ we change the variables $\phi,\xi\rightarrow \tilde{\phi},\tilde{\xi}$
\bea
\tilde \phi&=&(x-2p-5A,y,x+3p, x-2p-2A ,f,k,F,\Phi)\ , \label{tphidef}\\
\tilde \xi&=&\left(3\xi_1-\xi_3+\xi_4,\xi_2,-3\xi_1+2\xi_3-\xi_4,-3\xi_1+\xi_3-2\xi_4,\xi_5+\xi_6,\xi_5-\xi_6,\xi_7,\xi_8 \right)\ , \nn
\eea
while the inverse transformation is
\bea
\phi&=&(x,y,p,A,f,k,F,\Phi)\\&=&(\frac15 (-2\tphi_1+2\tphi_3+5\tphi_4),\tphi_2,\frac{1}{15} (2\tphi_1+3\tphi_3-5\tphi_4), \frac13 (-\tphi_1+\tphi_4) ,\tphi_5,\tphi_6,\tphi_7,\tphi_8)\ ,\nn \\
\xi&=&\left( \tfrac13(3\txi_1+\txi_3+\txi_4),\txi_2, \txi_1+\txi_3, -\txi_1- \txi_4,\tfrac12(\txi_5+\txi_6),\tfrac12(\txi_5-\txi_6),\txi_7,\txi_8\right)\ .\nn
\eea

The linearized equations of motion are (from now on we denote the perturbation part $\delta \tilde{\phi}$ simply by $\tilde{\phi}$)
 \bea
 \txi_1'&=&e^{-2x_0} \left(2P f_0-F_0(f_0-k_0)\right) \txi_1 \label{txi1eq} \\
 \txi_4'&=& -e^{-2x_0} \left(2P f_0-F_0(f_0-k_0)\right) \txi_1  \label{txi4eq} \\
\txi_5'&=&-\frac13 P e^{-2x_0} \txi_1 \label{txi5eq} \\
\txi_6'&=&-\txi_7-\frac13 e^{-2x_0} (P-F_0) \txi_1  \label{txi6eq} \\
\txi_7'&=&-\sinh(2y_0) \txi_5-\cosh(2y_0)\txi_6+\frac16 e^{-2x_0}(f_0-k_0) \txi_1 \label{txi7eq} \\
\label{txi8}
\txi_8'&=&(P e^{2y_0}-\sinh(2y_0)F_0) \txi_5 +(P e^{2y_0}-\cosh(2y_0)F_0) \txi_6+\frac12 (f_0-k_0) \txi_7\quad \ \ \ \  \label{txi8eq} \\
\txi_3'&=&3 e^{-2x_0-6p_0} \txi_3 +  \left(5 e^{-2x_0-6p_0} -e^{-2x_0}  (2P f_0-F_0 (f_0-k_0) \right) \txi_1 \label{txi3eq} \\
\txi_2'&=&\txi_2 \cosh y_0 +\frac13 \sinh y_0 (2 \txi_1 +\txi_3 +\txi_4) \non \\
&&\ \ \ \ \ \ \ \ +2\left((P e^{2y_0}-\cosh(2y_0)F_0) \txi_5 + (P e^{2y_0}-\sinh(2y_0)F_0) \txi_6\right)  \label{txi2eq}
   \eea
and
 \bea
  \tphi_8^{\prime} &=&- 4e^{-4(A_0+p_0)}\txi_8 \label{phi8peq} \\
  \tphi_2^{\prime} &=& - \cosh y_0\, \tphi_2 - 2 e^{-4(A_0+p_0)} \txi_2 \label{phi2peq}\\
 \tphi_3^{\prime} &=&  - 3 e^{-6p_0-2x_0}\tphi_3 - \sinh y_0\, \tphi_2 - \frac16 e^{-4(A_0+p_0)}  (9\txi_1+5\txi_3+2\txi_4)   \label{phi3peq}\\
  \tphi_1^{\prime} &=&  2 e^{-6p_0-2x_0} \tphi_3 -\sinh y_0 \tphi_2 + \frac{1}{6}e^{-4(A_0+p_0)}(\txi_1+3\txi_4)  \label{phi1peq}\\
 \tphi_5^{\prime} &=&   e^{2y_0}(F_0-2P) (2 \tphi_2 + \tphi_8) +e^{2y_0}\tphi_7
 - 2 e^{-4(A_0+p_0)+2(x_0+y_0)} (\txi_5 +\txi_6) \label{phi5peq}\\
 \tphi_6^{\prime}  &=&  e^{-2y_0} (F_0 (2\tphi_2 -\tphi_8)-\tphi_7)  - 2 e^{-4(A_0+p_0)+2(x_0-y_0)}  (\txi_5-\txi_6)  \label{phi6peq}\\
\label{eqtphi7}
 \tphi_7^{\prime} &=&   \half    \left( \tphi_5-\tphi_6 +(k_0-f_0)\tphi_8 \right) -2e^{-4(A_0 + p_0) +2x_0 } \txi_7 \label{phi7peq} \\
  \tphi_4^{\prime} &=& \frac{1}{5} e^{-2x_0} ( f_0 (2P-F_0) +k_0 F_0) (2 \tphi_1-2\tphi_3 -5 \tphi_4)  +\frac{1}{2} e^{-2x_0} (2P-F_0) \tphi_5    \non \ \ \ \ \ \ \  \\
 && +\frac{1}{2} e^{-2x_0} F_0  \tphi_6 +\frac{1}{2} e^{-2x_0}  (k_0 -f_0) \tphi_7 -\frac{1}{3} e^{-4(A_0+p_0)}\txi_1 \label{phi4peq}
 \eea

\subsection{Solving $\tilde{\xi}$ equations}
To solve the equations we follow the basic procedure outlined in \cite{Bena:2009xk}. First we solve the equations for $\txi_i$
and then deal with $\tphi_i$.
We notice that the equations (\ref{txi1eq}) can be integrated
\bea
\label{xi1sol}
\txi_1=X_1 h\ ,
\eea
where $X_1$ is some constant. The reason why the expression for $\txi_1$ is so simple will be explained below in Section 3.
The KS warp-factor $h$ is not an analytical function and since $\txi_1$ enters the equations for all other $\txi_i$ we won't be able to solve them explicitly.
Therefore we will present the solutions for $\txi_i$ assuming $X_1=0$ first and restore $X_1$ later.
The solutions parameterized by the constants $X_2,\dots,X_8$ are
\bea
\txi_1&=&0\ ,\\ \nn
\txi_4&=&X_4\ ,\\ \nn
\txi_5&=&X_5\ ,\\ \nn
\txi_6&=&{X_6\over 2\sinh(t)}+{X_7(\cosh(t)\sinh(t)-t)\over \sinh(t)}+{X_5 t\over \sinh(t)}\ ,\\ \nn
\txi_7&=&{X_6\cosh(t)\over \sinh(t)}-{X_7(\sinh(t)\cosh(t)^2+t\cosh(t)-2\sinh(t))\over \sinh(t)^2}+{X_5(t\cosh(t)-\sinh(t))\over \sinh(t)^2}\ ,\\ \nn
\txi_8&=&X_8+PX_5-{X_6 P(t\cosh(t)-\sinh(t))\over 2\sinh(t)^3} \\ \nn &&-{(X_7-X_5)P(t\cosh(t)-\sinh(t))(\cosh(t)\sinh(t)-t)\over \sinh(t)^3}\ ,\\ \nn
\txi_3&=&4X_3(\cosh(t)\sinh(t)-t)\ ,\\ \nn
\txi_2&=&2X_2\sinh(t) -{4X_3 t\cosh(t)\over 3} +{X_4\cosh ( t )\over 3} +
{X_5P(\sinh(t)^2-2\sinh(t)\cosh(t)t+t^2)\over \sinh(t)^3}\\&&-{X_6P(\cosh(t)\sinh(t)-t)\over 2\sinh(t)^3}+{X_7P(2\sinh(t)\cosh(t)^3 t-\sinh(t)^2-t^2)\over \sinh(t)^3}\ . \nn
\eea
Now, if we restore $X_1$ the solutions above will shift as follows $\txi_i\rightarrow \txi_i+X_1 \zeta_i$
\bea
\zeta_1&=&h\ ,\\ \nn
\zeta_4&=&-h\ ,\\ \nn
\zeta_5&=&-{16P\over 3}\int^t {dx\over (\cosh(t)\sinh(t)-t)^{2/3}}\ ,\\ \nn
\zeta_6&=&{(\cosh(t)\sinh(t)-t)\int^t {G(x) dx\over \sinh(x)}-\int^t{(\cosh(x)\sinh(x)-x)G(x) dx\over \sinh(x)}\over 2\sinh(t)}\ ,\\ \nn
\zeta_7&=&-\zeta_6'-{16P\over 3}{t\over \sinh(t)(\cosh(t)\sinh(t)-t)^{2/3}}\ ,\\ \nn
\zeta_8&=&-{P(t\cosh(t)-\sinh(t))\over \sinh(t)^2}\zeta_6\\ \nn &&-{16P\over 3}\int^t dx \left({P x (x\cosh(x)-\sinh(x))\over \sinh(x)^3 (\cosh(x)\sinh(x)-x)^{2/3}}\right.  \\  &&\nn+\left.{P(\sinh(x)(\cosh(x)^2+1)-2x\cosh(x))\over \sinh(x)^3}\int^x {dy\over (\cosh(y)\sinh(y)-y)^{2/3}}\right)\ ,\\ \nn
\zeta_3&=&-h+{2\over 3}(\cosh(t)\sinh(t)-t)\int^t {2\sinh(x)^2 h(x) dx\over (\cosh(x)\sinh(x)-x)^2}\ , \\ \nn
\zeta_2&=&\sinh(t)\int^t dx \left(-{2\over 9}{(\cosh(x)\sinh(x)-x)\over \sinh(x)^2}\int^x {2\sinh(y)^2 h(y) dy\over (\cosh(y)\sinh(y)-y)^2}\right.  \\ \nn
&& -\left.{32P^2\over 3}{t(\cosh(x)^2+1)-2\cosh(x)\sinh(x)\over \sinh(x)^4}\int^x {dy\over (\cosh(y)\sinh(y)-y)^{2/3}}\right.  \\ \nn
&& +\left.{2P(\sinh(x)(\cosh(x)^2+1)-2x\cosh(x))\over \sinh(x)^4}\int^x \zeta_6(y) dy \right)\ . \nn
\eea
where
\bea
\nn
G(t)={16P\over 3}\left({4 t\sinh(t)\over 3(\cosh(t)\sinh(t)-t)^{5/3}}+{2\cosh(t)\over \sinh(t)^2}\int^t {dx\over (\cosh(t)\sinh(t)-t)^{2/3}}\right)\ .
\eea

The zero energy condition $\xi_a{d\phi_0^a\over dt}=0$ implies
\bea
\nn
\txi_1{d(x+p-A)\over dt}+\txi_3{d(x/3+p)\over dt}+\txi_4{d(x/3-A)\over dt}+\txi_2{dy\over dt}+\txi_5{d(f+k)\over 2dt}+\txi_6{d(f-k)\over 2dt}+\txi_7{dF\over dt}\\
=-\frac 4 3 X_3+2X_2-PX_5+3P X_7=0\ .\quad\quad\quad\quad\quad\quad\quad  \label{ze}
\eea

\subsection{Solving $\tilde\phi$ equations}
Now we can proceed with the equations for $\tilde\phi$. Although we will find a general solution, we are primarily interested in the solution dual to the metastable state in the $SU(M(k+1)-p)\times SU(Mk-p)$ theory. The latter must have the same behavior at UV as the original KS solution (besides the shifted \Dt-charge). That will help us to fix seven integration constants by requiring that
all UV singular modes are turned off.  Besides, we will impose regularity at the IR of all modes except the one that is directly coupled to the \aDt-branes. That will fix another six integration constants. The boundary condition for the remaining singular mode will be fixed in Section 3.

The first equation for the perturbation of the dilaton $\tilde{\phi}_8$ is easy to solve
\bea
\label{tphi8}
\tilde{\phi}_8=Y_8+\int_t^\infty {64\over(\cosh(t)\sinh(t)-t)^{2/3}}\tilde{\xi}_8\ .
\eea
We choose the integration constant $Y_8=0$ such that $\tilde{\phi}_8\rightarrow 0$ at infinity i.e. there is no source for the operator ${\rm Tr}(F_1^2+F_2^2)$. The convergence at IR implies
\bea
X_8={P\over 6}(X_6-6X_5)\ .
\eea

The equation for $\tphi_2$ is also straightforward
\bea
\label{tphi2}
\tphi_2=-{32\over \sinh(t)}\int_0^t {\sinh(x)\txi_2(x) dx\over (\cosh(x)\sinh(x)-x)^{2/3}}+{Y_2\over \sinh(t)}\ .
\eea
Clearly to avoid $1/t$ singularity in the IR we put $Y_2=0$ and also require
\bea
\label{const:logt}
-16P^2\left({2\over 3}\right)^{1/3}X_1+{2\over 3}h_0X_1+PX_6-X_4=0
\eea
to avoid $\log(t)/t$ behavior at small $t$. In general $\tphi_2$  defined by (\ref{tphi2}) goes to zero as $e^{-t/3}$ at large $t$.
Given that the fluctuation $\tphi_2$ is dual to the operator ${\rm Tr}(\lambda_1\lambda_1+\lambda_2\lambda_2)$ of dimension $3$ we must require the leading asymptotic
$e^{-t/3}$ to vanish. As follows from (\ref{tphi2}) this would require the leading term of order $e^t$ in $\txi_2$ to vanish.
As there are two terms of the same order, $te^{t}$ and $e^{t}$, this implies two constraints
\bea
\label{const:dim3}
-{2\over 3}X_3+PX_7+\alpha X_1=0\ , \quad X_2+{X_4\over 6}+\beta X_1=0\ ,
\eea
where $(\alpha t+\beta)e^{t}$ is the UV asymptotic of $\zeta_2$. The coefficients $\alpha$ and $\beta$ can be calculated numerically. Their values are not important for what follows.

At first glance it looks surprising that the condition that there is no mass for ${\rm Tr}(\lambda_1\lambda_1+\lambda_2\lambda_2)$ leads to two, rather than to one condition.
In fact besides ${\rm Tr}(\lambda_1\lambda_1+\lambda_2\lambda_2)$ there is another operator ${\rm Tr}(\lambda_1\lambda_1-\lambda_2\lambda_2)$, also of dimension $3$.
Although these operators are independent, the corresponding modes mix in the KS case. The constraint (\ref{const:dim3}) ensures that both perturbations are absent.

At the next step we calculate $\tphi_3$
\bea
\tphi_3={1\over (\cosh(t)\sinh(t)-t)}\left[Y_3+\int_0^t \left({(\cosh(x)\sinh(x)-x)\over \sinh(x)}\tphi_2-\right. \right.\\ \left. \nn -\left.{8\over 3}(\cosh(x)\sinh(x)-x)^{1/3}(7\txi_1+5\txi_3+2X_4)\right)dx\right]\ .
\eea
Again, the integration constant $Y_3=0$ should be chosen to avoid the  $1/t^3$ singularity. To cancel $1/t$ behavior we must require
\bea
\label{const:1t}
X_6-16P\left({2\over 3}\right)^{1/3}X_1=0\ .
\eea
The $\log(t)/t$ term vanishes automatically at this point due to (\ref{const:logt}).
The leading $UV$ asymptotic of $\tphi_3\sim e^{2t/3}$ is due to $e^{2t}$ divergence of $\txi_3$. This mode is dual to the perturbation by the bottom component of the operator ${\rm Tr}W^2\bar W^2$ of dimension $6$ and
hence must be canceled by an appropriate choice of $X_3$
\bea
\label{const:dim6}
X_3+\gamma X_1=0\ .
\eea
The coefficient $\gamma$ can be  calculated numerically and, again, the exact value is not important for what follows.
This constraint assures the coefficient in front of the leading term $e^{2t}$ in $\txi_3$ vanishes.
In turn the first constraint (\ref{const:dim3})
implies that the coefficient in front of the leading term $e^{t}$ in $\txi_6,\txi_7$ vanishes as well.

The integration of $\tphi_1$ gives
\bea
\tphi_1&=&Y_1-\int^\infty_t dx \left({4\sinh(x)^2 \over 3(\cosh(x)\sinh(x)-x)}\tphi_3(x)+{\tphi_2(x)\over\sinh(x)}\right. \\
&&\nn \left.  + {8\over 3(\cosh(x)\sinh(x)-x)^{2/3}}(3X_4-2\txi_1)\right)\ .
\eea
We choose the integration constant $Y_1=0$ such that $\tphi_1$ goes to zero at infinity. In fact it is a gauge choice to keep $\tphi_1$ vanishing at infinity.
Finite value of $\tphi_1$ would only lead to a non-physical rescaling of  the warp-factor $A$.
As the IR  $\log(t)/t$ and $1/t$ behavior  vanishes due to (\ref{const:logt}) and (\ref{const:1t}) and the $\log(t)$ behavior vanishes due to the zero energy constraint (\ref{ze}),
we do not get any new constraints from regularity of $\tphi_1$ at small $t$.

The equations for $\tphi_{5},\tphi_{6},\tphi_{7}$ can be combined in such a way that the dependence on $\tphi_{5},\tphi_{6}$ drops and one gets a second-order differential equation for $\tphi_7$.
The corresponding solution depends on two integration constants
\bea\nn
\tphi_7&=&-{(\cosh(t)\sinh(t)-t)\int^\infty_t {G_7(x) dx\over \sinh(x)}+\int^t_0{(\cosh(x)\sinh(x)-x)G_7(x) dx\over \sinh(x)}\over 2\sinh(t)}\ \\ \nn
&&+{\tilde{Y}_7\over \sinh(t)}+{Y_7(\cosh(t)\sinh(t)-t)\over \sinh(t)},\\ \nn
G_7&=&{16P(t\cosh(t)-\sinh(t))\over \sinh(t)^2 (\cosh(t)\sinh(t)-t)^{2/3}}(4\txi_8-{2\over 3}\txi_1)\\&&-{2P(\sinh(t)(\cosh(t)^2+1)-2t\cosh(t))\over \sinh(t)^3}\tphi_2-2h'\txi_7\ .
\eea
The choice of $\tilde{Y}_7=0$ is dictated by the $1/t$ singularity in the IR.
In the case of general $\tphi_2$ the leading asymptotic of $G_7$ is $e^{-t/3}$ and therefore to cancel the $e^{t}$ growing term in $\tphi_7$ we required to put $Y_7=0$.
The corresponding perturbation of dimension $7$ is $\int d^2\theta W^2\bar W^2$.

The integration of $\tphi_5,\tphi_6$ is straightforward now
\bea\nn
\tphi_5&=& Y_5+\int_0^t dx \left({\cosh(x)-1\over \sinh(x)}\right)^2\left(-P\left(1+{x\over \sinh(x)}\right)(2\tphi_2+\tphi_8)+\tphi_7-2h (\txi_5+\txi_6)\right)\ ,\\ \nn
\tphi_6&=& {Y}_6   + \int_0^t dx\left({\cosh(x)+1\over \sinh(x)}\right)^2\left(  P\left(1-{x\over \sinh(x)}\right)(2\tphi_2-\tphi_8)-\tphi_7-2h (\txi_5-\txi_6)\right)\ .\\
\label{56}
\eea
As we started from three first order equations for $\tphi_{5,6,7}$  only three combination of $Y_5,Y_6,Y_7,\tilde{Y_7}$ are independent. There is a  constraint that expressed $Y_6$ through $Y_5$
\bea
\label{sing}
 Y_6=Y_5-{16\over 3}P \left({2\over 3}\right)^{1/3}h_0 X_1\ .
\eea
The fact that $Y_5=\tphi_5(0)$ is not equal to $Y_6=\tphi_5(0)$ leads to the divergent NS-NS field  $H^2\sim t^{-2}$ at small $t$. As is seen from (\ref{sing}) this singularity is always present for $X_1\neq 0$ i.e.
when the \aDt-branes are present \cite{Bena:2009xk}. In fact regular, although non-equal at the origin $\tphi_5$ and $\tphi_6$ corresponds to the IR regular mode of the linear system in question \cite{Berg:2006xy}. In particular this mode is normalizable in the IR.  There is also an IR singular mode with $\tphi_{5,6,7}$ diverging at the origin. That mode should not be turned on as there is no direct delta-function source for this mode coming from the probe action for the \aDt-branes.

As was mentioned above, absence of the perturbation by the operator ${\rm Tr}(\lambda_1\lambda_1-\lambda_2\lambda_2)$ is already covered by the constraint (\ref{const:dim3}). Indeed
in (\ref{const:dim6}) we required that the leading $e^{2t}$ asymptotic of $\tilde{\xi}_3$ vanishes. Therefore the only term potentially leading to $te^t$ behavior of $\txi_2$ (and correspondingly
to $te^{-t/3}$ behavior of $\tphi_2$) can come from the $e^t$ divergence of $\tilde{\xi}_6$. But the first constraint in (\ref{const:dim3}) is nothing but the requirement that $\tilde{\xi}_6$,
and hence $\txi_7$, does not have the leading $e^t$ term. As a result the fluctuation $\tphi_7$ given above goes to zero as $e^{-t}$, not as $e^{-t/3}$. This, together with the equation (\ref{eqtphi7})
 implies that $\tphi_5-\tphi_6$ approach zero at infinity also as $e^{-t}$. At the same time $\tphi_{5}+{\tphi_6}$ at infinity is some constant that is related to the net D3-charge. In general the  \Dt-charge is given by $-{M^2\over 4\pi P^2 }(f(2P-F)+kF)$.
Therefore shifting $f+k$ at infinity by a constant  while keeping $f-k,F$ the same  corresponds to changing the total D3-charge resulting from adding $\N$ \Dt\ or $\aN$ \aDt-branes
\bea
\label{bc}
f(r\rightarrow\infty)-f_{KS}(r\rightarrow\infty)=k(r\rightarrow\infty)-k_{KS}(r\rightarrow\infty)\rightarrow -{2\pi P\over M^2}(\N-\aN)\ \
\eea
The difference $f(r\rightarrow\infty)-f_{KS}(r\rightarrow\infty)$ is not quite the same as $\tilde{\phi}_5(t\rightarrow \infty)$
as the relation between the radial coordinate $r$ and the dimensionless coordinate $t$ involves the deformation parameter of conifold $\epsilon$: $r^3\sim \epsilon e^t$.
Since the deformation parameter could be different for the original KS background (we choose it to be $\epsilon=1$) and for the background perturbed by the \aDt-branes the condition (\ref{bc})
will determine $\epsilon$ as will be discussed in the next section.

The last remaining mode is
\bea
\label{tphi4}
\tphi_4&=&h^{-1}\left[Y_4-\int_t^\infty dx\left({2\over 5}h'(\tphi_1-\tphi_3)-{16X_1 h^2\over 3(\cosh(x)\sinh(x)-x)^{2/3}}\right.\right. \\ \nn
 &&+\left.\left.8P{(\sinh(x)+x)\tphi_5+(\sinh(x)-x)\tphi_6-2{x\cosh(x)-\sinh(x)\over\sinh(x)}\tphi_7\over \sinh(x)(\cosh(x)\sinh(x)-x)^{2/3}}\right)\right]\ .
\eea
Clearly the integration constant $Y_4$ must be put to zero. Otherwise the leading $e^{4t/3}$  behavior of $\tphi_4$ would correspond to the perturbation by the operator $\int d^2\theta d^2\bar\theta {\rm Tr}(W^2\bar W^2)$ of dimension $8$.
The IR behavior of $\tphi_4$ is singular
\bea
\label{deltahsing}
\delta h=2h\tphi_4={32\over 3}\left({3\over 2}\right)^{2/3} {X_1 h(0)^2-3P \tilde{\phi}_5(0)\over t}+..
\eea because of the delta-function at the origin associated with the presence of $\N$ \Dt\ or $\aN$ \aDt-branes \cite{Bena:2009xk}. We can relate the coefficient in front of the singular term in (\ref{deltahsing})
to the total number of $\N$ \Dt\ or $\aN$ \aDt-branes using the equations of motion which can be found in the next section
\bea
\label{deltah4}
{8\over 3\pi \alpha'^2}\left(X_1 h(0)^2-3PY_5\right)=\N+\aN\ .
\eea
This constrain can be understood as an equation for $Y_5$.

Thus we have fixed all integration constants except for $X_1$. We will do it in the next section.

\section{Solution for \aDt-branes}
As a warm-up we start with the solution for the \Dt-branes placed at the tip of the conifold i.e. put $\aN=0$ in the boundary condition (\ref{bc}). As a result we have $\tphi_5,\tphi_6$ shifted by a constant
\bea
\tphi_5=\tphi_6=-{2\pi P\over M^2}\N\ .
\eea
At the same time the \Dt-branes are mutually BPS with the background and therefore experience no force. This fixes $X_1$ (and hence all other constants) to be zero \cite{Bena:2009xk}. The only non-trivial quantity is the perturbation of the warp-factor
\bea
\label{deltah}
\delta h=2 h \tphi_4={4\pi \alpha'^2 \N }\int_t^\infty {dx\over (\cosh(x)\sinh(x)-x)^{2/3}}\ .
\eea
All other functions $\tphi_i$ for $i\neq 5,6,4$ are zero while $Y_5=\tphi_5(t)=\tphi_6(t)=-{2\pi P\over M^2}\N$ in agreement with (\ref{deltah4}).

The formalism used in this paper assumes the gauge choice $F_5=F_3\wedge B_2$. Hence the fields $f,k$ and their perturbations
$\tilde{\phi}_{5,6}$ control not only $B_2$ but also $F_5$. In fact we can make a gauge transformation shifting $f,k$ in $B_2$ by a constant while
keeping $F_5$ intact. It is convenient to choose a gauge such that the integral of $B_2\wedge F_3$ over $T^{1,1}$ vanishes at the origin. Then the value of $F_5$ at the origin defines the Page charge
\bea
Q_{D3}={1\over (4\pi^2\alpha')^2}\int_{T^{1,1}}F_5-B_2\wedge F_3\ .
\eea
For the small perturbation around the KS solution the Page charge is given by $Q_{D3}=-{M^2\over 2\pi P}Y_5$.
The Page charge is not gauge invariant. When the integral
\bea
\label{intS2}
{1\over 4\pi^2 \alpha'}\int_{S^2} B_2
\eea
becomes larger than $1$ one should make a gauge transformation shifting  (\ref{intS2}) back to zero.
This would change the Page charge by $M$. On the field theory side this corresponds to the Seiberg duality which changes ranks of the gauge groups. But the Page charge is quantized and mod $M$ conserved and therefore it counts (mod $M$) the number of \Dt\ or \aDt-branes placed at the origin.

If we start with the KS background and add $\N=lM$, $l\geq 1$ \Dt-branes the corresponding solution would be a gravity dual of the $l$-th mesonic branch of the $SU(M(k+1))\times SU(Mk)$ theory \cite{DKS}. To have the same behavior in the UV as the original KS solution the new solution must have different deformation parameter $\epsilon$. To find the relation between $\epsilon$ on different branches we can  look at the behavior of $k,f$ at infinity. For the original KS solution dual to the baryonic branch of moduli space we have
\bea
\label{bb}
f,k\rightarrow -P(t-1)=-P(\log r^3-\log \epsilon_0^2 -1)\ ,
\eea
while for the $l$-th mesonic branch
\bea
\label{mb}
f,k\rightarrow -P(t-1+{2\pi l\over M})=-P(\log r^3-\log \epsilon_l^2-1+{2\pi l\over M})\ .
\eea
The condition that (\ref{bb}) and (\ref{mb}) coincide gives
\bea
\epsilon_l^2=\epsilon_0^2 e^{2\pi l\over M}
\eea
in agreement with (\cite{gkp},\cite{remarks},\cite{DKS}).

Now we can proceed with the stack of $\aN=p$ \aDt-branes placed at the tip. First we require
\bea
\label{bcaDt}
f=k\rightarrow -P(\log r^3-\log \epsilon^2-1-{2\pi p\over M^2})\ .
\eea
at infinity or
\bea
\tilde{\phi}_5(t\rightarrow \infty)=\tilde{\phi}_6(t\rightarrow \infty)={2\pi P\over M^2}p+P\log \epsilon^2\ .
\eea
This will fix $\epsilon$ as a function of $X_i,Y_i$.   At the next step we fix $Y_5$ using (\ref{deltah}) with $\N=0, \aN=p$. Now we only need to fix $X_1$. An easy way to do that is to look at the force on probe $\N$ D3-branes placed at some location on the conifold in the presence of $\aN$ \aDt-branes located at the tip.
At first we would like to show that the resulting force has the expected form i.e. it is the same at the linear in $\N$ and $\aN$ order as the force on $\N$ probe \aDt-branes sitting at the tip produced by $N$ \Dt-branes sitting at some location on the conifold and affecting the background through the  backreaction  (\ref{deltah}).

To this end we consider the linearized equation for  $\Phi_\pm=e^{4\tilde A}\pm \alpha$ where the warp-factor $h=e^{-4\tilde A}$ (here we introduced $\tilde A=A-p-x/2$ to distinguish it from  $A$)
and $\alpha$ is the RR 4-form $C_4=\alpha\ dx^0\wedge..\wedge dx^3$.
The equations for $\Phi_\pm$ are \cite{gkp}
\bea
\label{phiplus}
\tilde\nabla^2 \Phi_+=e^{-4\tilde A}(\tilde\nabla \Phi_+)^2 +e^{8\tilde A}|\tilde G_+|^2+e^{8\tilde A}\N\kappa\  \delta({\rm location~ of~D3-branes})\ ,\\
\label{phiminus}
\tilde\nabla^2 \Phi_-=e^{-4\tilde A}(\tilde\nabla \Phi_-)^2 +e^{8\tilde A}|\tilde G_-|^2+e^{8\tilde A}\aN\kappa\  \delta({\rm location~ of~\overline{D}3-branes})\ .
\eea
Here $\kappa=2\kappa_{10}^2 2 T_3$ is some constant and we used the notations of \cite{gkp} for the ISD and AISD parts of the 3-form flux $G_\pm=(*_6\pm i)G_3$. In the case of the ISD solution, like the KS solution perturbed by some mobile \Dt-branes,  $\Phi_-=G_-=0$ and therefore at the linear order the equation for the warp-factor is
\bea
\tilde\nabla^2 h=-{1\over 2}|\tilde G_+|^2-{\kappa\over 2}\N \delta({\rm location~ of~D3-branes}).
\eea
Assuming we place \Dt's at some point $X_{D3}$ we have for the perturbation of $h$
\bea
\delta h(X)=-{\kappa \over 2}\N G(X,X_{D3})\ ,
\eea
where $G$ is the Green's function on the conifold \cite{Pufu:2010ie}.
Now treating the \aDt-branes as probes the corresponding \Dt-\aDt\  potential at the linear in $\N$, $\aN$ order is
\bea
\label{V1}
V_{D3-\overline{D}3}=2T_3 \aN h^{-1}(X_{\overline{D}3})={\kappa T_3 \N\aN \over h^2(X_{\overline{D}3})}G(X_{\overline{D}3},X_{D3})\ .
\eea

We can reproduce exactly the same answer (including the coefficient) in linear order
if we treat the \aDt-branes as sources changing the geometry. Then it follows from   (\ref{phiminus}) that at linear order
\bea
\tilde\nabla^2 \Phi_-=h^{-2}\aN\kappa\ \delta({\rm location~ of~\overline{D}3-branes})\ .
\eea

As a small remark, we emphasize that the mode $\Phi_-$ is the only one directly sourced by the delta-function coming from the \aDt-branes \cite{gkp}. That provides the IR boundary condition for $\tphi_{5}$\footnote{We thank the authors of \cite{Bena:2009xk} who pointed this out to us.}  (\ref{deltah4}) which follows from (\ref{phiplus}, \ref{phiminus}).

The solution for the linear mode of $\Phi_-$ is
\bea
\label{phiminusSol}
\Phi_-(X)={\kappa \aN \over  h^2(X_{\overline{D}3})}G(X,X_{\overline{D}3})\ ,
\eea
and the corresponding potential for the probe \Dt-branes at the linear in $\N$, $\aN$ level is
\bea
\label{V2}
V_{D3-\overline{D}3}=T_3\Phi_-={\kappa T_3 \N\aN \over h^2(X_{\overline{D}3})}G(X_{D3},X_{\overline{D}3})\ .
\eea
The results (\ref{V1}) and (\ref{V2}) are the same because of $G(X_{D3},X_{\overline{D}3})=G(X_{\overline{D}3},X_{D3})$.
Thus we have shown that the force on \Dt-branes induced by the \aDt-branes is the same as the force on the \aDt-branes induced by the \Dt-branes.
Clearly, the argument above is not based on smearing and is valid in the case of localized branes.

Now we can use the $SU(2)\times SU(2)$ invariant mode of (\ref{phiminusSol}) to fix $X_1$.
A direct calculation gives \cite{Bena:2009xk}
\bea
\Phi_-'={2\over 3}e^{-2x}\txi_1 \ .
\eea
This should match with the  $SU(2)\times SU(2)$ invariant mode of (\ref{phiminusSol}) with $\aN=p$
\bea
\Phi'_-={32 \tilde{X}_1\over 3 (\cosh(t)\sinh(t)-t)^{2/3}}\ , \quad \tilde{X}_1={12\pi  P^2 \aN\over h(0)^2 M^2}\ .
\eea This explains why the equation for $\txi_1$ can be integrated in the form (\ref{xi1sol}) and fixes  $X_1$ to be
\bea
\label{x1}
X_1={12\pi  P^2 p\over h(0)^2 M^2}\ .
\eea
This fixes all other integration constants $X_i,Y_i$ and hence determines in the unique way the lowest KK mode for the linearized solution describing $p$ \aDt-branes sitting at the tip of the deformed conifold.

The Page charge of the resulting solution is given  (mod $M$) by $Q=-{M^2\over 2\pi P}Y_5=-p$ in agreement with our expectations.

\section{Tension of \aDt-branes}
As a  check of our solution we will calculate the mass of the metastable state by calculating the ADM mass of the corresponding gravity solution found above. Our calculation will
closely follow a similar calculation carried out in \cite{dWKM}. The necessary ingredients are the extrinsic curvature of the eight-dimensional space $\mathcal K$ located at given time $x_0$ and infinite radius $r\rightarrow \infty$ and the norm of the time-like Killing vector $\mathcal N$. The resulting mass
\bea
E=-{1\over \kappa_{10}^2}\int {\mathcal N} {\mathcal K}\ ,
\eea
can be rewritten in the following form
\bea
\label{ADM}
\M=-{48\over (2\pi)^4\alpha'^4}e^{4A+4p}A'\ ,\quad  E=\int d^3 x\ \M\ .
\eea
This expression should be calculated at infinite $t$. It diverges as $e^{4t/3}\sim r^4$
and hence should be regularized, presumably by subtracting the tension $E_0$ of the supersymmetric configuration with the same UV asymptotic. Following \cite{BG} we subtract from (\ref{ADM}) the value of the superpotential at infinity
\bea
\label{ADMr}
\M_{\rm renorm}=-{48\over (2\pi)^4\alpha'^4}\left(e^{4A+4p}A'-{W\over 3}\right)\ .
\eea
This choice ensures that $\M_{\rm renorm}$ is zero for any supersymmetric solution that satisfies the BPS equations (\ref{susyeq}) i.e. for which all $\txi_i$ vanish. To calculate the mass for the configuration in question
we expand (\ref{ADMr}) at linear order and notice that only terms proportional to $\txi_i$ survive
\bea
\delta \M={8\over (2\pi)^4\alpha'^4}(\txi_1+\txi_4)\ .
\eea
Given that $\txi_1+\txi_4=X_4$ and using (\ref{const:logt}) and (\ref{const:1t}) and also the explicit expression for $X_1$ (\ref{x1}) we get for the tension of the meta-stable state at the linear in $\aN$ order
\bea
\label{ADMmass}
\M_{\rm renorm}={2 p T_3 \over h(0)} \ .
\eea
This result is in complete agreement with the probe approximation. Let us emphasize here that  (\ref{ADMmass}) depends on the IR integration constant $X_4$ which is sensitive to the IR boundary conditions. Hence our calculation provides a non-trivial check of the boundary conditions and the resulting solution.

\section{Gravity dual for the KS theory perturbed in UV}
As a by product of our analysis we obtained the general solution for the KS background perturbed by the $SU(2)\times SU(2)$ invariant, scalar, parity and ${\mathcal I}$-symmetry\footnote{The $\mathcal I$-symmetry acts by exchanging the $A$ and $B$ bifundamental fields of the KS theory accompanied by a charge conjugation. See e.g. \cite{Isymm}.} even operators  $W^2_{\pm}=W_1^2\pm W_2^2$ and $W^2_+\bar{W}_+^2$ with small coefficients. Some of these operators are irrelevant and the corresponding modes will destroy the AdS behavior at infinity. To avoid this the perturbation  at the UV cutoff scale $\Lambda$ should be small in the conventional units
\bea
\mathcal L_1\sim \varepsilon {{\mathcal O}_{\Delta}\over \Lambda^{\Delta-4}}\ ,
\eea
where dimensionless $\varepsilon\ll 1$.

Since a non-zero $X_1$ indicates presence of the \aDt-branes we put it to zero. This drastically simplifies the resulting solutions. The requirement that the solutions are regular in the IR implies $X_4=X_6=0$ and $X_8=-PX_5$.
At the same time $X_5$ can be expressed through $X_2,X_3,X_7$ with help of the zero energy condition (\ref{ze}).
Moreover we fix the gauge $Y_1=0$ to ensure the warp-factor behaves at infinity as in the unperturbed case.

Now we have seven integration constants left. They correspond: $X_2,X_7$ to bottom components of $W_\pm^2$;
$Y_8,Y_5$ to top components of $W_\pm^2$ i.e. coupling constants; $X_3$, $Y_7$ and $Y_4$ to bottom, medium and top components of $W^2_+\bar{W}_+^2$ correspondingly.

\subsection{KS with softly broken SUSY}
A particularly interesting example of the UV perturbation discussed above is a small  mass for gaugino bilinears $\lambda_{1,2}^2$ which softly breaks SUSY. This a relevant perturbation of field theory which corresponds to
 the fluctuation of metric $\tphi_2=\delta y$ and the three-form flux $\delta({1\over 2}(f-k)-F)={1\over 2}(\tphi_5-\tphi_6)-\tphi_7$.
The regularity at the IR together with the absence of perturbation by the irrelevant (and marginal) operators
in the UV results in two free parameters $X_2,X_7$
\bea
X_8=-PX_5=-(2X_2+3PX_7)\ ,\quad X_6=X_4=X_1=X_3=0\ ,\quad Y_i=0\ ,i\neq 5.
\eea
We should choose $Y_5=0$ as there is no \Dt\ or \aDt-branes at the origin. At the same time the deformation parameter $\epsilon$ should be chosen such that it compensates
$\tphi_5,\tphi_6$ at infinity.

Let us point out that a similar solution describing the KS theory perturbed by a dimension 3 operator was found in \cite{KuSo}. It corresponds to the one-parametric subfamily $PX_7=-X_2$.

\begin{center}
{\it Acknowledgments}
\end{center}
I am grateful to J. Maldacena for numerous discussions on the subject. I also benefited from the discussions with
O. Aharony, I. Bena, G. Giecold, M. $\rm Gra\tilde{n}a$, N. Halmagyi, S. Kachru, I. Klebanov, L. Martucci, S. Massai, S. Pufu, N. Seiberg, and H. Verlinde.
I am thankful to G. Giecold for reading the manuscript.
I gratefully acknowledge support from the Monell Foundation, the DOE grant DE-FG02-90ER40542,
and the Ministry of Education and Science of the Russian Federation under contract 14.740.11.0081.



\begin{thebibliography}{23}

\bibitem{ISS}
  K.~A.~Intriligator, N.~Seiberg, D.~Shih,
  ``Dynamical SUSY breaking in meta-stable vacua,''
  JHEP {\bf 0604}, 021 (2006).
  [hep-th/0602239].

\bibitem{KPV}
  S.~Kachru, J.~Pearson and H.~L.~Verlinde,
  ``Brane/Flux Annihilation and the String Dual of a Non-Supersymmetric Field
  Theory,''
  JHEP {\bf 0206}, 021 (2002)
  [arXiv:hep-th/0112197].

\bibitem{KS}
  I.~R.~Klebanov and M.~J.~Strassler,
  ``Supergravity and a confining gauge theory: Duality cascades and
  chiSB-resolution of naked singularities,''
  JHEP {\bf 0008}, 052 (2000)
  [arXiv:hep-th/0007191].


\bibitem{dWKV}
  O.~DeWolfe, S.~Kachru and H.~L.~Verlinde,
  ``The giant inflaton,''
  JHEP {\bf 0405}, 017 (2004)
  [arXiv:hep-th/0403123].

\bibitem{PS}
  J.~Polchinski and M.~J.~Strassler,
  ``The string dual of a confining four-dimensional gauge theory,''
  arXiv:hep-th/0003136.


\bibitem{dWKM}
  O.~DeWolfe, S.~Kachru and M.~Mulligan,
  ``A Gravity Dual of Metastable Dynamical Supersymmetry Breaking,''
  Phys.\ Rev.\  D {\bf 77}, 065011 (2008)
  [arXiv:0801.1520 [hep-th]].


\bibitem{KT}
  I.~R.~Klebanov and A.~A.~Tseytlin,
  ``Gravity Duals of Supersymmetric SU(N) x SU(N+M) Gauge Theories,''
  Nucl.\ Phys.\  B {\bf 578}, 123 (2000)
  [arXiv:hep-th/0002159].

\bibitem{mGSS}
  P.~McGuirk, G.~Shiu and Y.~Sumitomo,
  ``Non-supersymmetric infrared perturbations to the warped deformed
  conifold,''
  Nucl.\ Phys.\  B {\bf 842}, 383 (2010)
  [arXiv:0910.4581 [hep-th]].


\bibitem{Bena:2009xk}
  I.~Bena, M.~Grana and N.~Halmagyi,
  ``On the Existence of Meta-stable Vacua in Klebanov-Strassler,''
  JHEP {\bf 1009}, 087 (2010)
  [arXiv:0912.3519 [hep-th]].

\bibitem{Bena:2010ze}
  I.~Bena, G.~Giecold, M.~Grana and N.~Halmagyi,
  ``On The Inflaton Potential From Antibranes in Warped Throats,''
  arXiv:1011.2626 [hep-th].

\bibitem{Bena:2011hz}
  I.~Bena, G.~Giecold, M.~Grana, N.~Halmagyi, S.~Massai,
  ``On Metastable Vacua and the Warped Deformed Conifold: Analytic Results,''
  [arXiv:1102.2403 [hep-th]].


\bibitem{BG}
  V.~Borokhov and S.~S.~Gubser,
  f``Non-supersymmetric deformations of the dual of a confining gauge  theory,''
  JHEP {\bf 0305}, 034 (2003)
  [arXiv:hep-th/0206098].

\bibitem{KuSo}
  S.~Kuperstein and J.~Sonnenschein,
  ``Analytic non-supersymmetric background dual of a confining gauge theory
  and the corresponding plane wave theory of hadrons,''
  JHEP {\bf 0402}, 015 (2004)
  [arXiv:hep-th/0309011].





\bibitem{gkp}
  S.~B.~Giddings, S.~Kachru and J.~Polchinski,
  ``Hierarchies from fluxes in string compactifications,''
  Phys.\ Rev.\  D {\bf 66}, 106006 (2002)
  [arXiv:hep-th/0105097].

\bibitem{Berg:2006xy}
  M.~Berg, M.~Haack and W.~Mueck,
  ``Glueballs vs. gluinoballs: Fluctuation spectra in non-AdS/non-CFT,''
  Nucl.\ Phys.\  B {\bf 789}, 1 (2008)
  [arXiv:hep-th/0612224].

\bibitem{DKS}
  A.~Dymarsky, I.~R.~Klebanov, N.~Seiberg,
  ``On the moduli space of the cascading SU(M+p) x SU(p) gauge theory,''
  JHEP {\bf 0601}, 155 (2006).
  [hep-th/0511254].

\bibitem{remarks}
  C.~P.~Herzog, I.~R.~Klebanov and P.~Ouyang,
  ``Remarks on the warped deformed conifold,''
  arXiv:hep-th/0108101.

\bibitem{Pufu:2010ie}
  S.~S.~Pufu, I.~R.~Klebanov, T.~Klose and J.~Lin,
  ``Green's Functions and Non-Singlet Glueballs on Deformed Conifolds,''
  arXiv:1009.2763 [hep-th].

\bibitem{Isymm}
  A.~Dymarsky, D.~Melnikov and A.~Solovyov,
  ``I-odd sector of the Klebanov-Strassler theory,''
  JHEP {\bf 0905}, 105 (2009)
  [arXiv:0810.5666 [hep-th]].



\end{thebibliography}
\end{document}